\documentclass[journal=jacsat,manuscript=article]{achemso}

\usepackage{amsmath,amssymb}
\usepackage[version=3]{mhchem}
\usepackage{graphicx}
\usepackage[ansinew]{inputenc}
\usepackage{hyperref,color}

\author{Fei Gao}
\affiliation{Department of Physics, Technical University of Denmark, DK-2800 Kongens Lyngby, Denmark}

\author{Rodrigo E. Mench\'on}
\affiliation{Donostia International Physics Center (DIPC), 20018 Donostia-San Sebasti\'an, Spain}

\author{Aran Garcia-Lekue}
\affiliation[1]{Donostia International Physics Center (DIPC), 20018 Donostia-San Sebasti\'an, Spain}
\alsoaffiliation[2]{IKERBASQUE, Basque Foundation for Science, 48013 Bilbao, Spain}
\email{wmbgalea@ehu.es}

\author{Mads\ Brandbyge}
\affiliation{Department of Physics, Technical University of Denmark, DK-2800 Kongens Lyngby, Denmark}
\email{mabr@dtu.dk}

\title[An \textsf{achemso} demo]
  {Tunable spin and transport in porphyrin-graphene nanoribbon hybrids}

\date{\today}

\begin{document}

\begin{abstract}
Recently, porphyrin units have been attached to graphene nanoribbons (Por-GNR) enabling a multitude of possible structures. Here we report first principles calculations of two prototypical, experimentally feasible, Por-GNR hybrids, one of which displays a small band gap relevant for its use as electrode in a device. Embedding a Fe atom in the porphyrin causes spin polarization with a spin ground state $S=1$. We employ density functional theory and nonequilibrium Green's function transport calculations to examine a 2-terminal setup involving one Fe-Por-GNR between two metal-free, small band gap, Por-GNR electrodes. The coupling between the Fe-$d$ and GNR band states results in a Fano anti-resonance feature in the spin transport close to the Fermi energy. This feature makes transport highly sensitive to the Fe spin state. We demonstrate how mechanical strain  or chemical adsorption on the Fe give rise to a spin-crossover to $S=1$ and $S=0$, respectively, directly reflected in a change in transport. Our theoretical results provide a clue for the on-surface synthesis of Por-GNRs hybrids, which can open a new avenue for carbon-based spintronics and chemical sensing.

\end{abstract}
\maketitle

\section{Introduction}

Graphene nanoribbons (GNRs) with extensive $\pi$-delocalized electrons have attracted attention due to their electronic properties like width-dependent band gaps, edge states and long spin relaxation time, turning these one-dimensional (1D) materials into promising building blocks for nanoelectronic and spintronic devices in the carbon family \cite{GNR-1,GNR-2, GNR-3,long-spin,de2022carbon}. One of the most significant pigments in nature, porphyrin, exhibits tunable spin properties in its conjugated arrays, depending on the central metal ion and the surrounding ligand field \cite{porphyrin-1,porphyrin-arrays,jurow2010porphyrins-2}. Therefore, it would be a highly appealing strategy to synthesize porphyrin-graphene nanoribbon (Por-GNR) hybrids with well-ordered atomic arrangements and possibilities of tailoring the band gap, topological phases, and detecting magnetic signals in transport. 

During the last decade, on-surface synthesis has become a powerful technique to form atomically precise nanostructures by linking small precursor molecules \cite{on-surface-1, on-surface-2,on-surface-3,on-surface-4} or porphyrin building blocks \cite{porphyrin-arrays-1, porphyrin-arrays-3} in a bottom-up approach. This makes the fabrication of quasi 1D Por-GNR hybrids feasible, avoiding problems such as random molecular placement and metal clusters formation \cite{clustering1,clustering2}. Recently, several research groups have attempted to expand the synthetic 1D complexes including porphyrin cores in different ways\cite{1D-1,mallada2021surface-1D}. The structures considered so far, both in experimental and theoretical works, have mainly been porphyrin oligomers/polymers \cite{por-ploymers}, or porphyrin nanotapes \cite{Porphyrin-Nanotapes}, which lack the GNRs segments as a backbone. However, Mateo {\it et al.}, have synthesized structures with two metal-free ($H_\text{2}$) Pors connected by a short GNR segment \cite{Por-GNRs}. This naturally poses the question of how electronic transport takes places in GNRs with incorporated Pors, and especially spin transport for Pors with magnetic centers.

In this letter, we propose two Por-GNRs hybrids which might be experimentally feasible using two existing carbon-based precursor molecules \cite{synthesis-of-GNR-2010, synthesis-of-GNR} and a porphyrin center. Combining such  molecular units gives rise to the straight hybrid1 and ``S-shape'' hybrid2 structures shown in Fig.~\ref{fig:Geometry}. Our first principles calculations reveal that hybrid2 has a small electronic band-gap ($\sim 0.1-0.2$ eV), which is highly desirable for potential spintronic devices. Embedding an iron atom in the porphyrin center in both 1D nanostructures gives rise to spin polarization with a spin ground state of $S = 1$. Employing the nonequilibrium Green's function (NEGF) formalism, we consider a 2-terminal device setup including one Fe-hybrid2 linked between two H$_\text{2}$-hybrid2. The coupling between the electronic states of the GNR segments and those states with matching symmetry in the Fe-porphyrin center, leads to spin-polarized Fano anti-resonances close to the Fermi energy in the transmission. A switching of the spin-state to $S = 2$ can be achieved by mechanical strain, while the adsorption of a CO molecule on top of the Fe center fully quenches the magnetism. Our work highlights the potential of Por-GNRs as a highly tunable and flexible platform for spintronics and sensing applications.

\section{Results and discussion}

\begin{figure}
\includegraphics[scale=0.69]{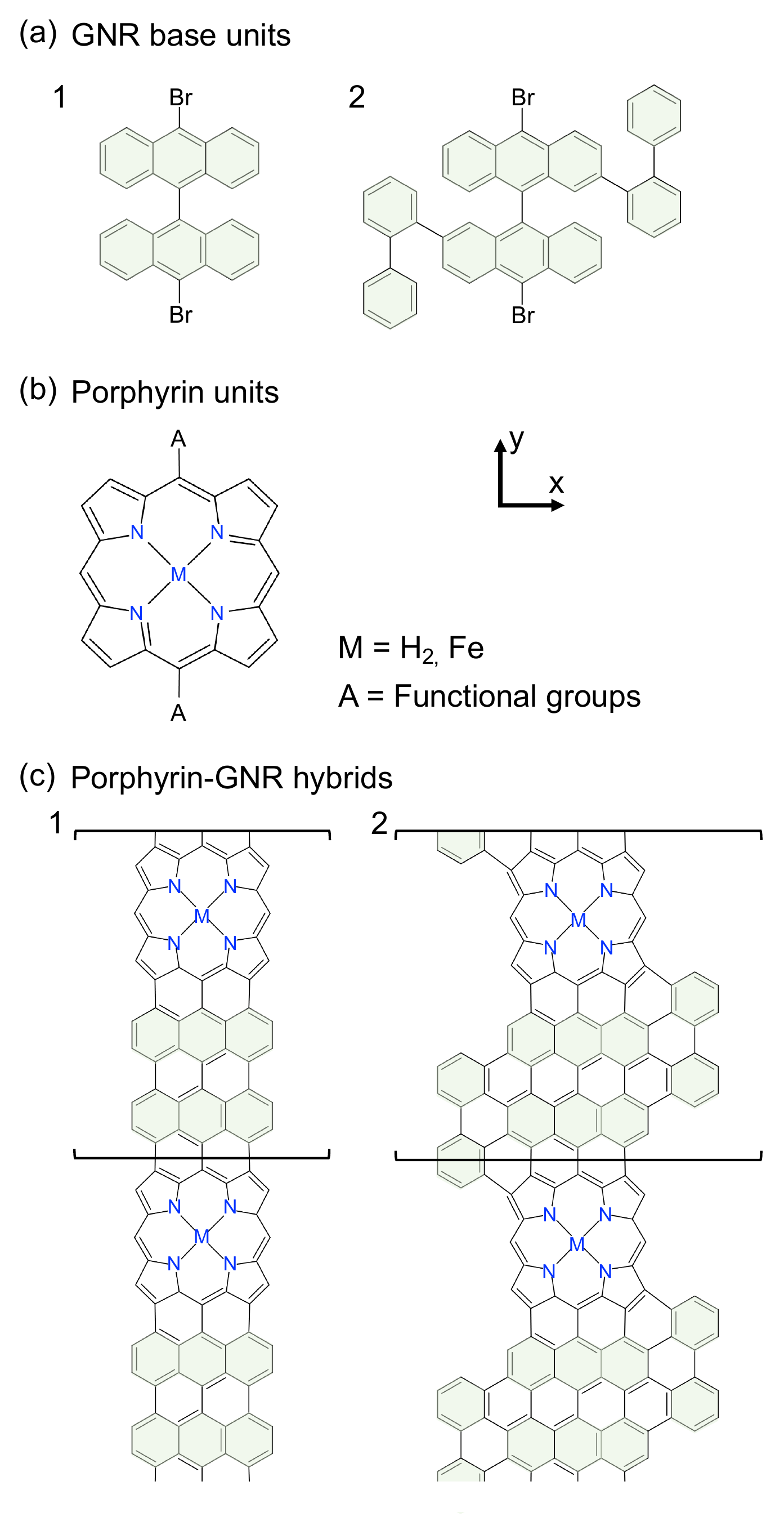}
\caption{\label{fig:Geometry}(a) Schematic diagram of two carbon-base precursor molecules (GNR-base units). (b) Porphyrin (Por) building block. Here, M = $H_\text{2}$ and Fe, and A stands for functional groups. (c) Schematic diagram of 1D infinite Por-GNR hybrids: 1 as straight-shape Por-GRN hybrid1 and 2 as S-shape Por-GNR hybrid2.} 
\end{figure}

Figure ~\ref{fig:Geometry}a and b show two existing carbon-based precursor units (1,2) and the porphyrin building block, respectively, with M representing either a metal-free ($H_\text{2}$) or a metallized (e.g. with a Fe atom) unit. The precursors we select here have already been synthesized and successfully used to grow atomically precise 7-AGNRs (unit 1) \cite{7-AGNRs} and 13-AGNRs \cite{13-AGNRs} (unit 2) on surface. In Fig~\ref{fig:Geometry}c we show how the combination of Por with units 1 or 2 can give rise to a quasi 1D Por-GNR structure which may be repeated periodically or joined by additional ``pristine'' 1 or 2 units, respectively. Moreover, the different precursors produce GNR segments of distinct morphology, resulting in a
straight-shaped hybrid1 and a ``S-shaped'' hybrid2. The previous successful bottom-up synthesis of GNRs and Por-GNR connections indicates that these proposed systems should be feasible. 

\begin{figure}
\includegraphics[scale=0.6]{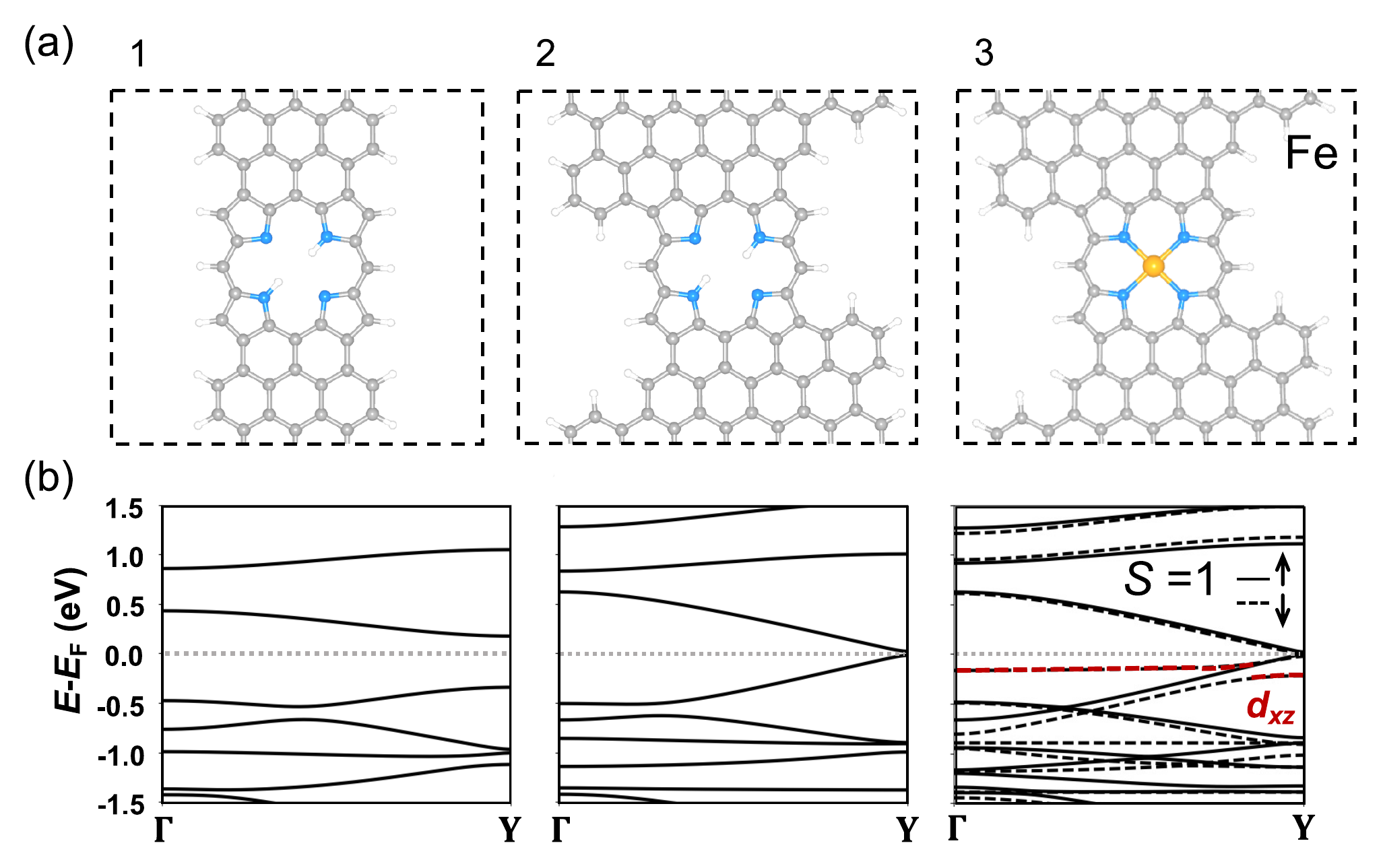}
\caption{\label{fig:band}(a) The optimized structures of 1D Por-GNR hybrids. The gray, blue, white and yellow balls represent C, N, H and Fe atoms, respectively. (b) The corresponding band structure. For Fe-Por-GNR hybrid2, the black solid and dash lines represent the spin up and spin down states, respectively, and the red dash line is labelled the spin down state of Fe-$d_{{xz}}$. Here, the gray dotted line is the Fermi level.}
\end{figure}

The DFT optimized structures in Fig. ~\ref{fig:band}a are planar for the two periodically repeated $H_\text{2}$-Por-GNR hybrids, and both are non-spin polarized, as seen in the band structures in Fig. ~\ref{fig:band}b. Besides, the bands shown in the left and middle panels in Fig. ~\ref{fig:band}b indicate that different edge conformations lead to different frontier bands and, importantly, reveal a band gap closing in hybrid2 at the Y-point (zone-boundary). Although this result depends on the DFT exchange functional employed, the trend is the same: the straight hybrid1 has a larger gap of 0.75 eV for PBE and 0.9 eV for HSE06, whereas the ``S-shape'' hybrid2 displays a small electronic gap of 0.1 eV for PBE and 0.25 eV for HSE06. This band-gap ``closing'' is  likely to be caused by the presence of wider GNR segments in hybrid2, as the  band gap of the pristine 13-AGNRs is around 1 eV larger than that of 7-AGNRs \cite{13-AGNRs}. The small band gap exhibited by hybrid 2 would be highly useful for potential applications in quantum transport nanodevices. 
To gain a deeper insight into the properties of the Por-GNR hybrids, we calculate the ${\mathbb{Z}_\text{2}}$ topological invariant for 
non-metallized hybrid1 and hybrid 2, using the supercells shown in left and middle panels of Fig ~\ref{fig:band}a. We obtain ${\mathbb{Z}_\text{2}}$ = 1 for both structures, indicating that they are both
in a topologically non-trivial phase and that 
localized end states at the interface to vacuum are be expected. Moreover, as shown in Fig.~S1, we do not observe any symmetry inversion when inspecting the wavefunctions of the conduction and valence bands at $\Gamma$ and Y for both Por-GNR structures, also confirming that such two hybrids belong to the same topological family.

In an attempt to realize future spintronics in these novel nanostructures, we embed an iron atom in the porphyrin center in hybrid2 (Fe-hybrid2). This gives rise to spin polarization, the ground state being the $S=1$ solution. The atomic structure of $D_{4h}$ symmetry remains planar and the four equivalent Fe-N bonds have a length of 2~\AA. For comparison, we also introduce a Fe atom into hybrid1 and the relaxed flat structure has the same occupation of Fe-3$d$ orbitals as Fe-hybrid2 (see Fig. S2),  with a slightly decrease Fe-N bond length of 1.97~\AA. The ground state of the isolated iron tetraphenyl porphyrin (FeTPP), which contains the same central macrocycle, has a  $S=1$ ground state
($^3\!A_{2g}$) and the occupancy of the 3$d$ shell is $(d_{x^2-y^2})^2(d_{z^2})^2(d_{xz})^1(d_{yz})^1$, where the last two orbitals are degenerate as a consequence of the molecular symmetry \cite{liao2002electronic, FeTPP_Gr}. However, for the ground state of the hybrid2 we obtain the $^3\!E_{g}$ electronic configuration, in which the 3$d$ occupation is $(d_{x^2-y^2})^2(d_{z^2})(d_{xz})^2(d_{yz})^1$, despite the total spin moment being still 2 $\mu_\text{B}$. It is the breaking of symmetry and the coupling of the GNR band states to the Fe-porphyrin states with matching symmetry which results in this change. The right panel in Fig. ~\ref{fig:band}b illustrates the electronic band structure of Fe-Por-GNR hybrid2, with the occupied spin down state of Fe-$d_{{xz}}$ (marked in red) very close to the Fermi level (at $\approx$ -0.15 eV).
In addition, we obtain a meta-stable state with the occupation corresponding to $^3\!A_{2g}$, at a modest energy expense of 0.1 eV (see Fig. S3). For the isolated Fe-porphyrin molecule, the energy difference between $^3\!A_{2g}$ and $^3\!E_{g}$ is always very small $\sim 10$~meV \cite{FeTPP_Gr}. This implies that the spin state of the central iron atom can be stabilized by coupling with the GNRs segments, revealing one of the distinct advantages of the proposed 1D hybrids. Indeed, the spin-multiplet energetics of nearly degenerate spin-states is a difficulty for this research area at least from the theoretical perspective \cite{multi,multi-1,FeTPP_Gr}.

\begin{figure}
\includegraphics[scale=0.5]{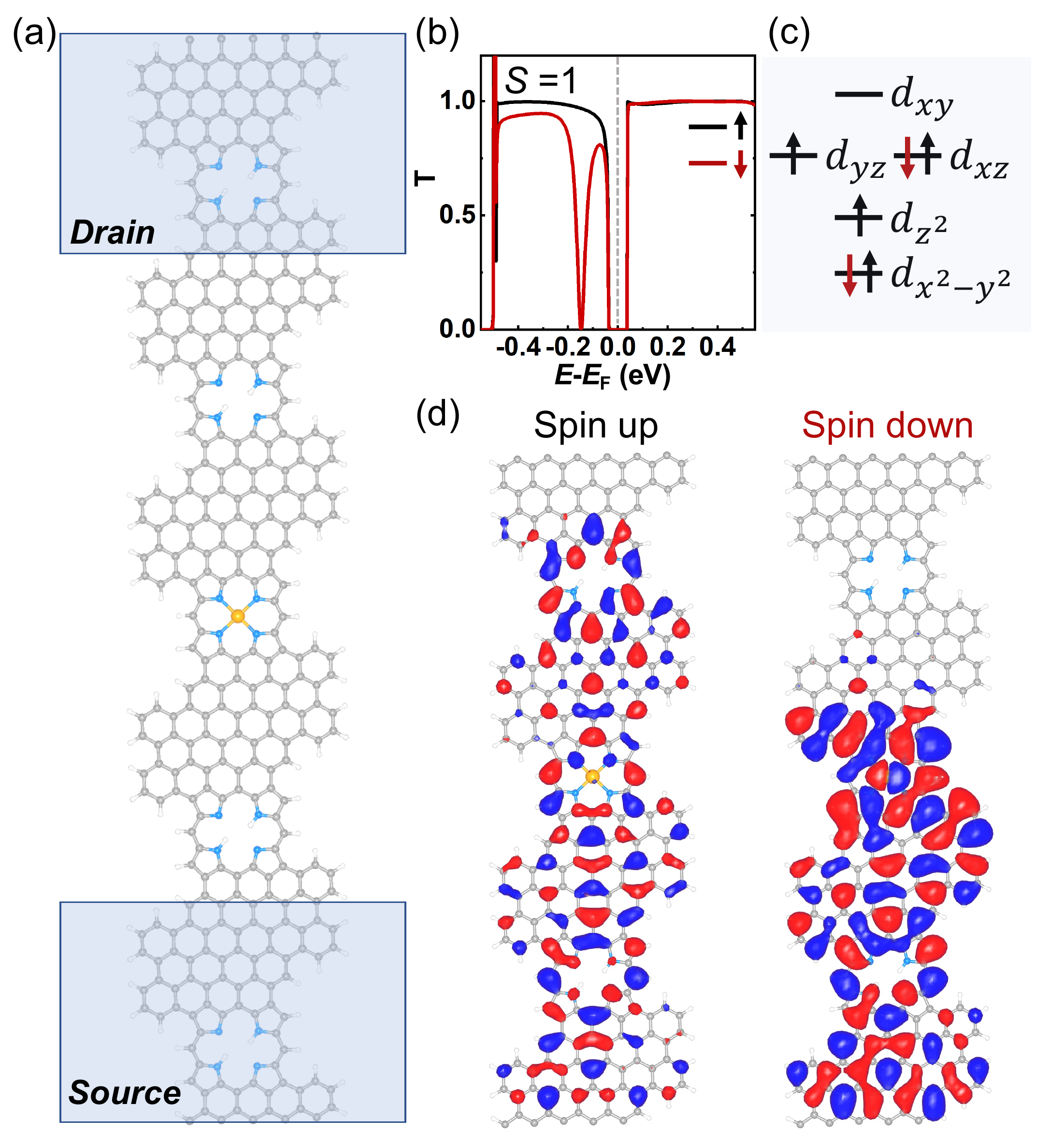}
\caption{\label{fig:transport} (a) Two-terminal transport setup. The shaded areas (Source and Drain) are the electrodes. (b) Zero-bias transmission. The black and red lines represent the spin up and spin down states, respectively. (c) Occupation of Fe $d$-orbitals in the $S=1$ state in the device. (d) The real part of the (source-to-drain) eigenchannel scattering states at -0.15 eV. Both spin up and spin down channels in the scattering region are shown. Red and blue clouds indicate the positive and negative sign of the wavefunction, respectively. Isosurfaces with values of $\pm 0.01$ e/$\text{bohr}^\text{3}$ are shown.}
\end{figure}

Motivated by this understanding, we now consider the electronic transport properties of the 1D Por-GNR hybrid2. We create a two-terminal setup in order to access the spin-dependent transport of one isolated Fe-hybrid2 unit sandwiched between two H$_\text{2}$-hybrid2 in the device region, the latter being repeated in the semi-infinite left/right directions acting as electrodes, as shown in Fig. ~\ref{fig:transport}a. Since we employ pristine H$_\text{2}$-hybrid2 as electrodes, there is a small gap at the Fermi level in the zero-bias transmission function (see Fig. ~\ref{fig:transport}b). Figure ~\ref{fig:transport}c describes the 3$d$ level occupation of the Fe atom in the device region, which is the same as for the Fe atom in the periodical Fe-Por-GNR hybrid2 ($^3\!E_{g}$). Interestingly, a zero-bias transmission dip occurs for spin down channel (red line in Fig. ~\ref{fig:transport}b) at -0.15 eV related to the 3$d_{{xz}}$ orbital of Fe atom. The interference between waves involving the 3$d$ state on the Fe atom, and waves directly transmitted in band states not involving the Fe states, yields a Fano resonance. We further plot the real part of the source-to-drain eigenchannel scattering state \cite{eigenchannel1, eigenchannel2} for the energy corresponding to the dip (-0.15 eV) in Fig. ~\ref{fig:transport}d. The wavefunction for the spin up state goes through the device region, while it is fully reflected and therefore vanishing on the top side for the spin down state. From the sign and shape of the wavefunctions we can identify the orbitals involved in the transport($d_{xz}$ on Fe).

\begin{figure*}
\includegraphics[scale=0.5]{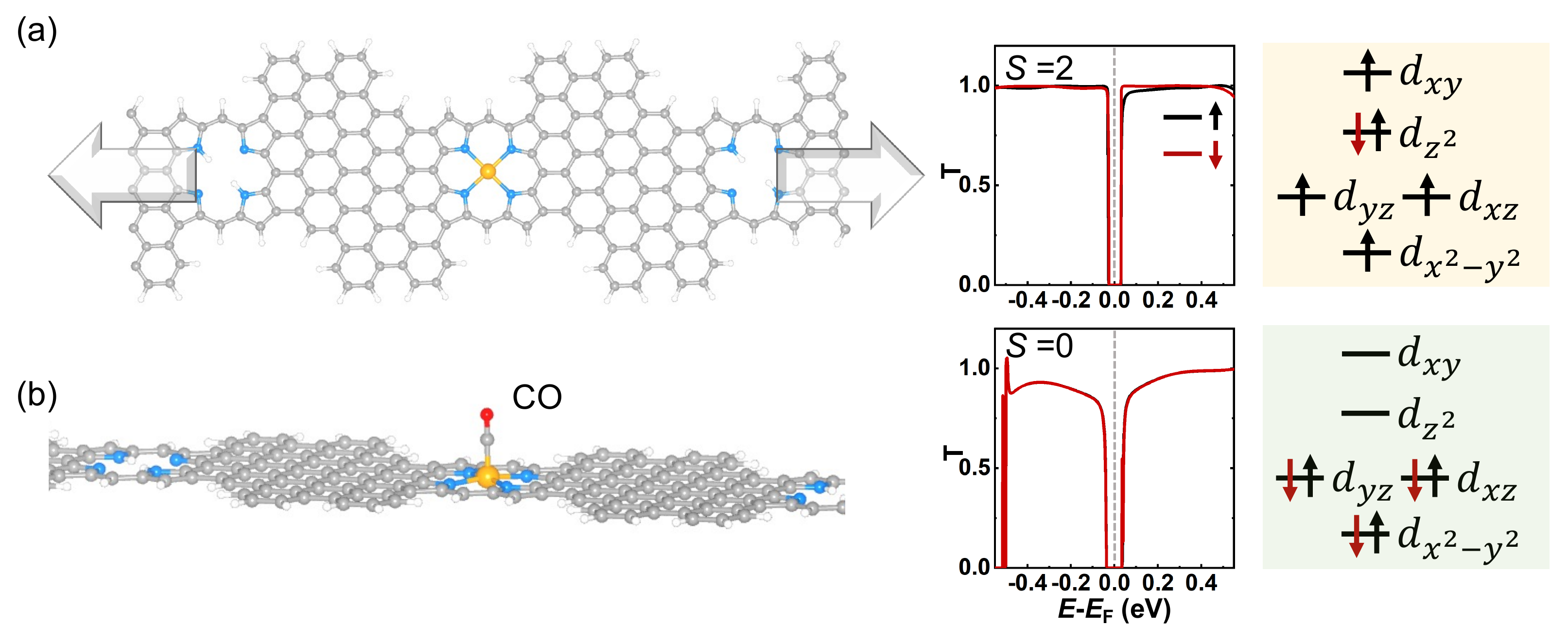}
\caption{\label{fig:application}(a) Schematic diagram of strain effect on the Fe atom in 1D Por-GNR hybrid2 device. (b) Schematic diagram of CO on top of the Fe atom in Fe-Por-GNR hybrid2 device. The corresponding zero-bias transmission and occupation of Fe $d$-orbitals are shown in the middle and right panels.} 
\end{figure*}

The Fe porphyrin-based nanomaterials, presenting an open $d$-shell, allow for various possible electronic configurations and different spin-multiples. This can result in a spin-crossover (SCO) behavior controlled by the environment and external stimuli. In Fig.~\ref{fig:application} we demonstrate how we may manipulate the spin state of the 1D device discussed above, either by mechanical strain along the GNR backbones, or by chemical adsorption of a CO molecule on the Fe center. Figure ~\ref{fig:application}a shows the mechanically strained SCO device and the resulting spin transition from $S=1$ to $S=2$ obtained by applied 3$\%$ uniaxial strain. 
The four Fe-N bond lengths in the stretched flat geometry increase from 2 to 2.10~\AA, leading to the variation of the Fe-$d$ occupation and spin state, as shown in the right panel in Fig. ~\ref{fig:application}a. This also results in a drastic change in transport, specially in the valence band where the Fano feature disappears, when compared to the unstrained case in Fig.~\ref{fig:transport}b. The strain results in almost equal spin up and spin down transmissions in the energy range from -0.5 to 0.5 eV. Thus our proposed Por-GNR hybrid2 device is a good candidate for a mechanically driven spin filter device.

Another way to change the spin state in the same device setup is the adsorption of a CO molecule on the top of the Fe atom, with a binding energy of 2.03 eV. In this case the total magnetic moment varies from 2 to 0 $\mu_\text{B}$ due to the strong covalent binding. In detail, the Fe-CO bond length is 1.73~\AA\, and the C-O distance is 1.16~\AA (see Fig. ~\ref{fig:application}b). Interestingly, the Fe-N bond length remains 2~\AA\, and there is also little vertical displacement of the Fe atom. The Mulliken charge analysis shows that upon adsorption only 0.2 electrons are transferred from CO to the Fe-3$d$ system, and the density of states projected on the 3$d$ orbitals of the Fe atom displays a stronger crystal-field splitting (see Fig. S4). Here the change of the crystal field from a square planar to a square pyramidal symmetry is responsible for the spin transition. This indicates that our proposed Por-GNR hybrid2 device would be a promising material for next-generation chemical sensing \cite{Fe-N-C}.

\section{Conclusion}

We have performed first principles DFT-NEGF calculations to study the electronic and transport properties of two proposed metal-free Por-GNRs hybrids based on already synthesized precursors. The 1D dimensional ``S-shaped'' Por-GNR system (hybrid2) reveals a small band gap, and is therefore interesting as a device electrode candidate. Embedding an iron atom in the porphyrin center causes spin polarization with  $^3\!E_{g}$ ground state, due to the coupling of the GNR band states to the Fe-porphyrin. A feasible 2-terminal transport setup including one Fe-hybrid2 is considered, and a Fano anti-resonance dip appears in the zero-bias transmission of the spin down channel close to the Fermi energy. We demonstrate how this prominent feature and the resulting transmission around the Fermi level can be manipulated by external stimuli that change the spin state, such as an applied mechanical strain ($S=2$ @ 3\% strain), or the adsorption of a CO molecule which quench the magnetic moment ($S=0$). In light of potential applications in spintronics, it is worth exploring possible magnetic properties including the surface effect (Kondo resonance), topological quantum phase\cite{topological-quantum-phases}, spin-spin interaction \cite{porphyrin-arrays-2} and the possibility of the 2D structures \cite{2Dmagnetism} in the novel, spin-hosting Por-GNRs hybrids, which will be pursued in future investigations.

\section{Computational procedures}

We employed the SIESTA/TransSIESTA code, GGA-PBE \cite{PBE} for exchange-correlation, and a DZP basis-set \cite{siesta, brandbyge2002density, transiesta}. The energy cutoff of 400 Ry was used to define the real-space grid. All the calculations were carried out with a low electronic temperature of 50 K and spin polarization was also included. The k-point mesh of $1\times6\times1$ was used and a 25~{\AA} thick vacuum layer for the slab model was introduced. Importantly, all atoms were allowed to relax until the forces on each atom are smaller than 0.01 eV/~\AA. Moreover, we considered the mean-field correction of the Hubbard $U$ = 3 eV for Fe 3$d$ orbitals \cite{Dudarev1998}. The computational parameters used in the equilibrium calculations were checked carefully for convergence and reproduced accurately the trends obtained by the Vienna ab initio simulation package (VASP) \cite{vasp} with PBE \cite{PBE} and HSE06 \cite{HSE06} functionals. Subsequently, physical quantities like the density of states, transmission, current and spin density were extracted using SISL \cite{sisl}.
Zak phases were obtained from the electronic contributions to the macroscopic electric polarization values calculated with the SIESTA code.\cite{siesta,polarizationKingSmith1993}

\begin{acknowledgement}
The authors thank Aitor Mugarza and Diego Pe\~{n}a for fruitful discussions and highly valuable input. This project received funding from the EU Horizon 2020 under Grant No. 766726. The authors gratefully acknowledge financial support from Grant PID2019-107338RB-C66 funded by MCIN/AEI/10.13039/501100011033, the European Union (EU) H2020 program through the FET Open project SPRING (Grant Agreement No. 863098), the Dpto. Educaci\'on Gobierno Vasco (Grant No. PIBA-2020-1-0014) and the Programa Red Guipuzcoana de Ciencia, Tecnolog\'iae Innovaci\'on 2021 (Grant Nr. 2021-CIEN-000070-01. Gipuzkoa Next). The authors also acknowledge the financial support received from the IKUR Strategy under the collaboration agreement between Ikerbasque Foundation and DIPC on behalf of the Department of Education of the Basque Government.
\end{acknowledgement}

\bibliography{main} 

\end{document}